\documentclass[a4paper, amsfonts, amssymb, amsmath, reprint, showkeys, nofootinbib, onecolumn, preprint]{revtex4-1}
\usepackage[utf8]{inputenc}
\usepackage[colorinlistoftodos, color=green!40, prependcaption]{todonotes}

\usepackage{graphicx}    
\usepackage{kotex} 
\usepackage{xcolor} 

\usepackage{soul}

\setstcolor{red}

\begin{document}
\title{Revisiting the Gamow Factor of Reactions on Light Nuclei}
\author{Eunseok Hwang}
\affiliation{Department of Physics and OMEG Institute, Soongsil University, Seoul 06978, Republic of Korea}

\author{Heamin Ko}
\email{Corresponding author: h@soongsil.ac.kr}
\affiliation{Department of Physics and OMEG Institute, Soongsil University, Seoul 06978, Republic of Korea}

\author{Kyoungsu Heo}
\affiliation{Department of Physics and OMEG Institute, Soongsil University, Seoul 06978, Republic of Korea}

\author{Myung-Ki Cheoun}
\affiliation{Department of Physics and OMEG Institute, Soongsil University, Seoul 06978, Republic of Korea}

\author{Dukjae Jang}
\email{Corresponding author: djjang2@ibs.re.kr}
\affiliation{Center for Relativistic Laser Science, Institute for Basic Science (IBS), Gwangju 61005, Korea}

\date{\today}

\begin{abstract}
 This study provides an improved understanding of the penetration probabilities (PPs) in nuclear reactions of light nuclei by correcting the assumptions used in the conventional Gamow factor. The Gamow factor effectively describes the PP in nuclear reactions based on two assumptions: low particle energy than the Coulomb barrier and neglecting the dependence of nuclear interaction potential. However, we find that the assumptions are not valid for light nuclei. As a result of a calculation that excludes the assumptions, we obtain the PP that depends on the nuclear interaction potential depth for the light nuclei. For the potential depth fitted by the experimental fusion cross-section, we present that PPs of light nuclei (D+D, D+T, D+$^3$He, p+D, p+$^6$Li, and p+$^7$Li) become higher than the conventional one near the Coulomb barrier. We also discuss the implications of the modified PP, such as changes in the Gamow peak energy, which determine the measurement of energy range of the nuclear cross-section in experiments, and the electron screening effect.
\end{abstract}

\maketitle

%
%
\section{Introduction}
One of the ultimate goals of nuclear astrophysics is to elucidate the origin of elements in the universe. For this purpose, astrophysical processes such as $r-$, $p-$, and $s-$processes \cite{Kappeler_1989,1957RvMP...29..547B,ARNOULD200797,1994ARA&A..32..153M,THIELEMANN2011346} have been well established, and recently, the $rp-$ \cite{1998PhR...294..167S} and $\nu-$induced processes \cite{2006PhRvL..96n2502F,1990ApJ...356..272W,2019ApJ...872..164K,2022ApJ...937..116K} have been developed to supplement the conventional astrophysical processes. Such studies on nucleosynthesis are commonly evaluated by the network calculation, in which thermonuclear reaction rate plays a crucial role as the main building block.

For the two-body reaction of $i + j \to k + l$, the thermonuclear reaction rate per a pair of  particls is generally given as\footnote{We adopt the natural unit of $\hbar=c=k_B \equiv 1$ for all equations in this paper.}
\begin{eqnarray} \label{eq:reaction_rate}
    \langle \sigma v \rangle_{ij}
    &=& \sqrt{\frac{8}{\pi \mu_{ij} T^3}} \int E e^{-E/T} \sigma(E)dE,
\end{eqnarray}
where $\sigma(E)$ is the cross-section for the given nuclear reaction, $v$ is the relative velocity between species $i$ and $j$, $\mu_{ij}$ is the reduced mass of $i$ and $j$, $T$ is the temperature, and $E = \mu_{ij}v^2/2$ is the relative energy of $i + j$ particles. In the typical temperature range of nucleosynthesis environments, the thermonuclear reaction rate in Eq.\,(\ref{eq:reaction_rate}) is characterized by the cross-section in the energy region from sub-keV to a few MeV. However, it still remains a challenge to measure the low energy cross-section of nuclear reactions as these energy regions are near or below the Coulomb barrier.

The low energy cross-section can be explained by factorizing $\sigma(E)$ into the contributions by nuclear and Coulomb interaction as follows:
\begin{equation}
	\sigma(E)= \sigma_{\rm nuc}(E) \hat{P}(E)
\end{equation}
where $\sigma_{\rm nuc}(E)$ represents the cross-section by nuclear interaction, and $\hat{P}(E)$ is the penetration probability (PP) for the Coulomb barrier. For the nuclear interaction part, the $\sigma_{\rm nuc}(E)$ can be expressed as $\sigma_{\rm nuc}(E) = S(E)/E$ using the astrophysical S-factor, $S(E)$. For the Coulomb interaction part, the $\hat{P}(E)$ is given as
\begin{equation}
	\hat{P}(E)= e^{-2\pi \eta},
\label{eq:gamow_factor}	
\end{equation}
where $\eta$ is the Sommerfeld parameter defined as $\eta \equiv Z_i Z_j e^2/v$ with nuclear charges $eZ_i$ and $eZ_j$. The PP in Eq.\,(\ref{eq:gamow_factor}) is known as the Gamow factor, which effectively describes the probability of the charged nuclei penetrating the given Coulomb barrier. Then, using $S(E)$ and $\hat{P}(E)$, the nuclear cross-section is rewritten as
\begin{eqnarray}
\sigma(E) = \frac{S(E)}{E} e^{- 2\pi \eta}, 
\end{eqnarray}
When a low-lying resonance is absent, the $S(E)$ is monotonic at the low energy region. 
To determine the $S(E)$ precisely at these low energies, the R-matrix theory \cite{1958RvMP...30..257L,2010RPPh...73c6301D} has been utilized, providing a more precise evaluation of $S(E)$. Therefore, even with the lack of low energy cross-section data, the nuclear cross-section for nucleosynthesis can be evaluated using the extrapolated $S(E)$ and the Gamow factor, a widely adopted method in nuclear astrophysics \cite{2022PhRvC.105a4625O,2019ApJ...872...75D}.

However, it is questionable whether the conventional Gamow factor is valid for reactions of light nuclei. The Gamow factor in Eq.\,(\ref{eq:gamow_factor}) is derived from the Wentzel-Kramers-Brillouin (WKB) approximation in the low energy limit, with the assumption that the particle energy is much lower than the Coulomb barrier. This WKB approximation results in the Gamow factor that depends only on the nuclear charge and energy. While the low energy assumption is suitable for reactions involving heavy nuclides, the Coulomb barriers for the light nuclei could be comparable to the thermal kinetic energy in specific astrophysical environments. For instance, in Big-Bang nucleosynthesis or supernova explosions, temperature can reach a few MeV, which are not significantly lower than the Coulomb barriers $(V_1=Z_1 Z_2 e^2/R_0$ where $R_0= r_0 (A_1^{1/3}+A_2^{1/3})$ with $r_0$ in \cite{2019NuPhA.986...98S}) tabulated in Table \ref{tab:couloumb_bariirer}. This implies that the assumption used to derive the conventional Gamow factor may not be proper. 
\begin{table}[h]
    \begin{tabular}{c | c}
        \hline \hline        
        \textbf{reaction} & \textbf{Couloumb barrier (MeV)} \\
        \hline            
            ${\rm D+D}$      & 0.206 \\ 
            ${\rm D+T}$      & 0.338 \\ 
            ${\rm D+ ^3He}$  & 0.320 \\ 
            ${\rm p+D}$      & 0.541 \\ 
            ${\rm p+^6Li}$   & 1.300 \\ 
            ${\rm p+^7Li}$   & 1.115 \\ 
        \hline \hline 
    \end{tabular}
    \caption{Couloumb barrier for light nuclei reactions. }
    \label{tab:couloumb_bariirer}
\end{table}

Hence, this paper aims to numerically investigate the penetration probability (PP) for charged nuclei without the assumptions adopted for the Gamow factor. Our findings present a PP that depends on the nuclear interaction potential depth for the reactions listed in Table \ref{tab:couloumb_bariirer}. We compare the fully calculated PP to the conventional Gamow factor, using a potential depth fitted by the fusion cross-section data. Furthermore, we discuss the implication of the modified PP on the Gamow peak energy, the measurement of nuclear cross-section in experiment, and the electron screening effect.

This paper is organized as follows: In section II, we introduce the modified PP that depends on potential depth, evaluating the validity of assumptions for the conventional Gamow factor. In section III, we explore the modified Gamow peak energy resulting from the change in the PP. In Section IV, we discuss possible change in  the electron screening effect due to the modified PP. Section V provides the conclusion of this paper.

%
%
\section{MODIFIED GAMOW FACTOR}
In a reaction between two charged nuclides, we should take into account two kinds of potentials for nuclear and Coulomb interactions. The description of the nuclear interaction potential is model-dependent, so various effective potentials are used to describe the low energy reactions, such as optical potential \cite{Satchler1983}, folding potential \cite{1979PhR....55..183S}, Aky\"uz-Winther potential \cite{Aky_Win_1979}, and so on. However, it is challenging to obtain the precise bare potential for light nuclei from these models \cite{1962AnPhy..19..287F}. In this study, we adopt the square well potential commonly used for the reactions of light nuclei, which has been shown to describe the fusion cross-section to be consistent with the result based on the complex potentials \cite{1970PhRvC...2.2041M,2000PhRvC..61b4610L,2019PhRvC..99f5808O}. Using the square well potential with the real potential, we derive the penetration probability. On the other hand, the Coulomb potential is exactly given as a shape of $1/r$, so we can treat it as a summation of infinitesimal square barriers.

For the potentials of the square well and infinitesimal barrier, we can derive the PP of the reaction between $i$ and $j$, respectively \cite{2015nps..book.....I},  
\begin{eqnarray}
    \label{eq:P_W}
    \hat{P}_{W}(E,V_0) &=& \frac{4\sqrt{(E+V_0)E}} {\left[ \sqrt{(E+V_0)}+\sqrt{E} \right]^2}, \\[12pt]
     \label{P_B}
    \hat{P}_{B}(E, V_1) &=& \frac{4E} {\left[4E +\left( E+V_1+\frac{E^2}{V_1-E} \right) \sinh^2 (\sqrt{2\mu_{ij} (V_1-E)} \Delta) \right]},
\end{eqnarray}
where $\hat{P}_{W}(E, V_0)$ and $\hat{P}_{B}(E, V_1)$ represent the PP for the square well and infinitesimal square barrier, respectively. The $\hat{P}_{W}$ depends on the particle energy $E$ and the depth of the well $V_0$, and the $\hat{P}_{B}(E, V_1)$ depends on the height of the barrier $V_1$, $\mu_{ij}$, and width of the square barrier potential $\Delta$ as well as $E$. Then, combining both potentials, we can obtain the PP as \cite{2015nps..book.....I}
\begin{eqnarray}
    \label{eq:P_WB}
    \hat{P}_{WB}(E, V_0, V_1) &=&  4 \sqrt{E(E+V_0)}  
        \left[2E+V_0+2\sqrt{E(E+V_0)} \right. \nonumber \\ [12pt]
    && \left. +\left( E+V_0+V_1+\frac{E(E+V_0)}{V_1-E} \right) \sinh^2 (\sqrt{2\mu_{ij} (V_1-E)} \Delta) \right]^{-1}.
\end{eqnarray}
To derive the conventional Gamow factor, two approximations are used based on the following assumptions. First, the particle energy is much lower than the Coulomb barrier, i.e., $E \ll V_1$, which leads Eq.\,(\ref{eq:P_WB}) to
\begin{eqnarray}
	\hat{P}_{WB}(E, V_0, V_1) \approx \frac{16 \sqrt{E(E+V_0)}(V_1-E)}{V_1(V_0+V_1)} \exp \left( {-2 \sqrt{2\mu_{ij} (V_1-E)}\Delta} \right ).
	\label{eq:gc}
\end{eqnarray}
Second, the coefficient of the exponential factor in Eq.\,(\ref{eq:gc}) is assumed to be an order of unity for reasonable physical values of $E$, $V_0$, and $V_1$, which results in
\begin{eqnarray}
\hat{P}_{WB}(E,V_1) \approx \exp \left( {-2 \sqrt{2\mu_{ij} (V_1-E)} \Delta} \right ).
\label{eq:gc2}
\end{eqnarray}
We note that the last approximation removes the dependence of the nuclear interaction potential $V_0$ on PP. Since the PP includes only the infinitesimal barrier in Eq.\,(\ref{eq:gc2}), the total PP is obtained by accumulating each PP from the classical turning point ($R_C$) to the radius of the potential well ($R_0$). This product can be expressed as the summation of each exponent. For the infinitesimal width in the summation, $\Delta \ll 1$, we can convert the total PP to the following integration form: 
\begin{eqnarray}
    \hat{P}_G(E) \approx \exp \left[ -2 \int^{R_C}_{R_0} \sqrt{2\mu_{ij} \left(\frac{Z_i Z_j e^2}{r}-E\right)}dr  \right],
    \label{Gamow_factor}
\end{eqnarray}
The leading term of the integration in Eq.\,(\ref{Gamow_factor}) is equivalent to Eq.\,(\ref{eq:gamow_factor}), the conventional Gamow factor.

Although the assumptions used in Eqs.\,(\ref{eq:gc}) and (\ref{eq:gc2}) are mostly valid, especially for heavy nuclei, the Coulomb barriers are not significantly high for reactions of light nuclei as shown in Table \ref{tab:couloumb_bariirer}. Thus, we modify the conventional PP using Eqs.\,(\ref{eq:P_W}) and (\ref{P_B}) without the aforementioned two approximations, whose formula is written as follows:
\begin{eqnarray}
    \label{eq:modified_Gamow_factor}
    \hat{P}_{mod}(E, V_0,V_1) = \hat{P}_{WB,1}(E, V_0, V_1) \prod^n_{m=2} \hat{P}_{B,m}(E, V_{1,m}),
\end{eqnarray}
where $\hat{P}_{mod}(E, V_0,V_1)$ represents the modified PP and the subscript $m$ indicates $m$-th PP for the infinitesimal barrier from the innermost barrier. Each barrier in Eq.\,(\ref{eq:modified_Gamow_factor}) depends on the height of $m$-th barrier, $V_{1,m}$, at each position, $R_{m}$. These quantities are respectively given as:
\begin{eqnarray}
    \label{R_i}
    R_m &=& \left(\frac{R_c-R_0}{n}\right)(m-1)+R_0,\\[12pt]
    \label{eq:V_1,i}
    V_{1,m} &=& \frac{Z_1 Z_2 e^2}{R_m}.
\end{eqnarray}
For the innermost barrier, i.e., $m=1$, the PP includes both barrier and square well potential. Since we take into account the potential depth, the general form of $\hat{P}_{mod}(E, V_0,V_1)$ in Eq.\,(\ref{eq:modified_Gamow_factor}) depends on nuclear interaction potential depth $V_0$ and $V_1$, and $E$.

\begin{figure}[h] 
    \centering
    \includegraphics[width=16 cm]{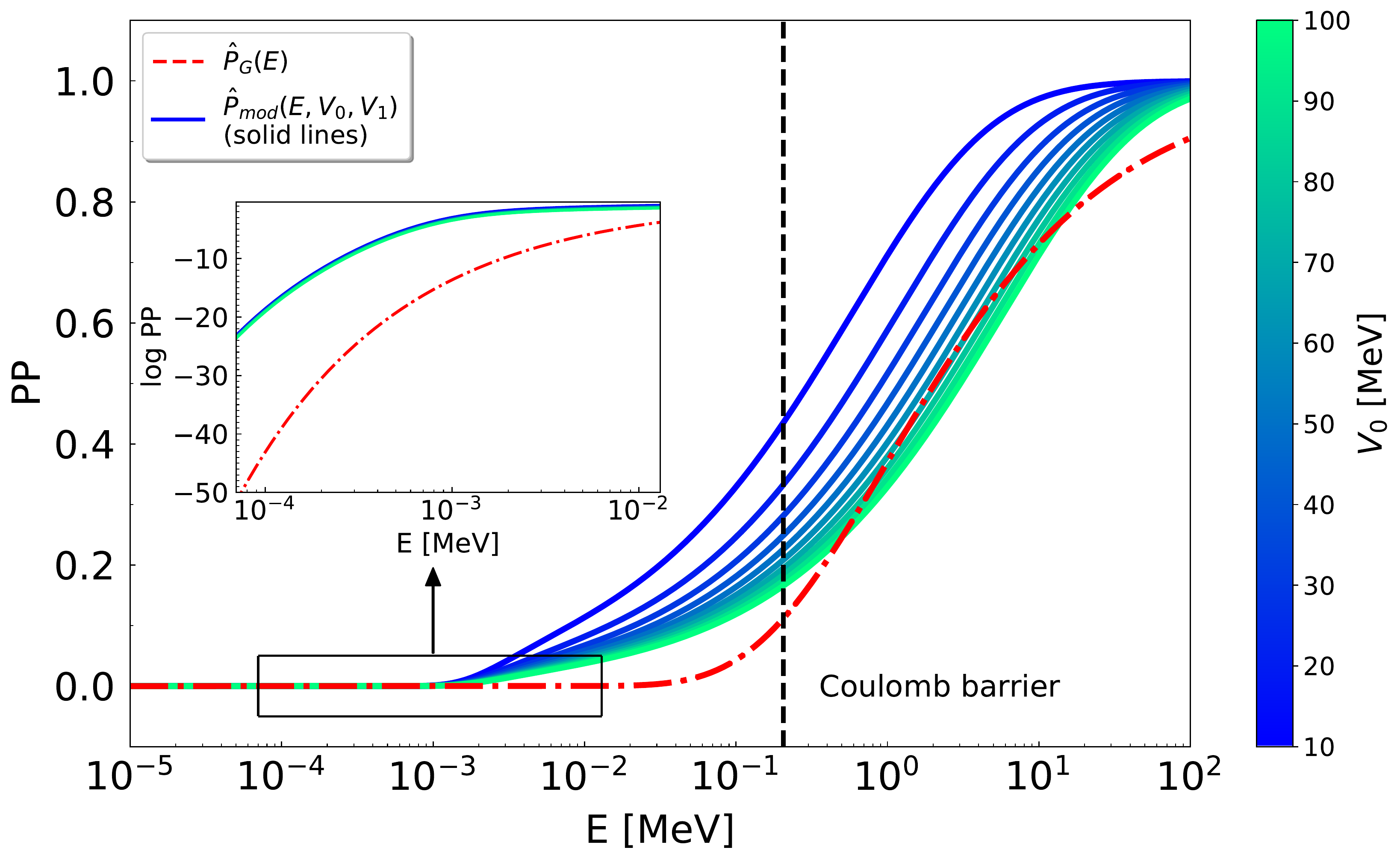}
    \caption{The PP for the D+D reaction as a function of $E$ and $V_0$. The solid and dot-dashed lines represent $\hat{P}_{G}(E)$ and $\hat{P}_{mod}(E, V_0,V_1)$, respectively. The color bar on the right indicates value of  $V_0$ for $\hat{P}_{mod}(E, V_0,V_1)$. The black-vertical dashed line marks the Coulomb barrier,  $V_1 = 0.206\,{\rm MeV}$ for the D+D reaction. A subplot provides magnified plot within the energy range from 0.7 keV to 13 keV.}
    \label{fig:depth}
\end{figure}

 Figure \ref{fig:depth} shows PP for the D+D reaction as a function of $E$ and $V_0$. Both $\hat{P}_{G}(E)$ and $\hat{P}_{mod}(E, V_0,V_1)$ increase with particle energy $E$, indicating that the higher the energy of a particle, the higher the probability of penetrating the Coulomb barrier. However, the difference between $\hat{P}_{G}(E)$ and $\hat{P}_{mod}(E, V_0,V_1)$ is the dependence on $V_0$. While $\hat{P}_{G}(E)$ does not depend on $V_0$, $\hat{P}_{mod}(E, V_0,V_1)$ increases as $V_0$ decreases. This is because $\hat{P}_{mod}(E,V_0,V_1)$ is similar to $\hat{P}_W(E,V_0)$, that decreases by increase of $V_0$ near the Coulomb barrier where $E \simeq V_1$.

\begin{figure}[h] 
    \centering
    \includegraphics[width= 15 cm]{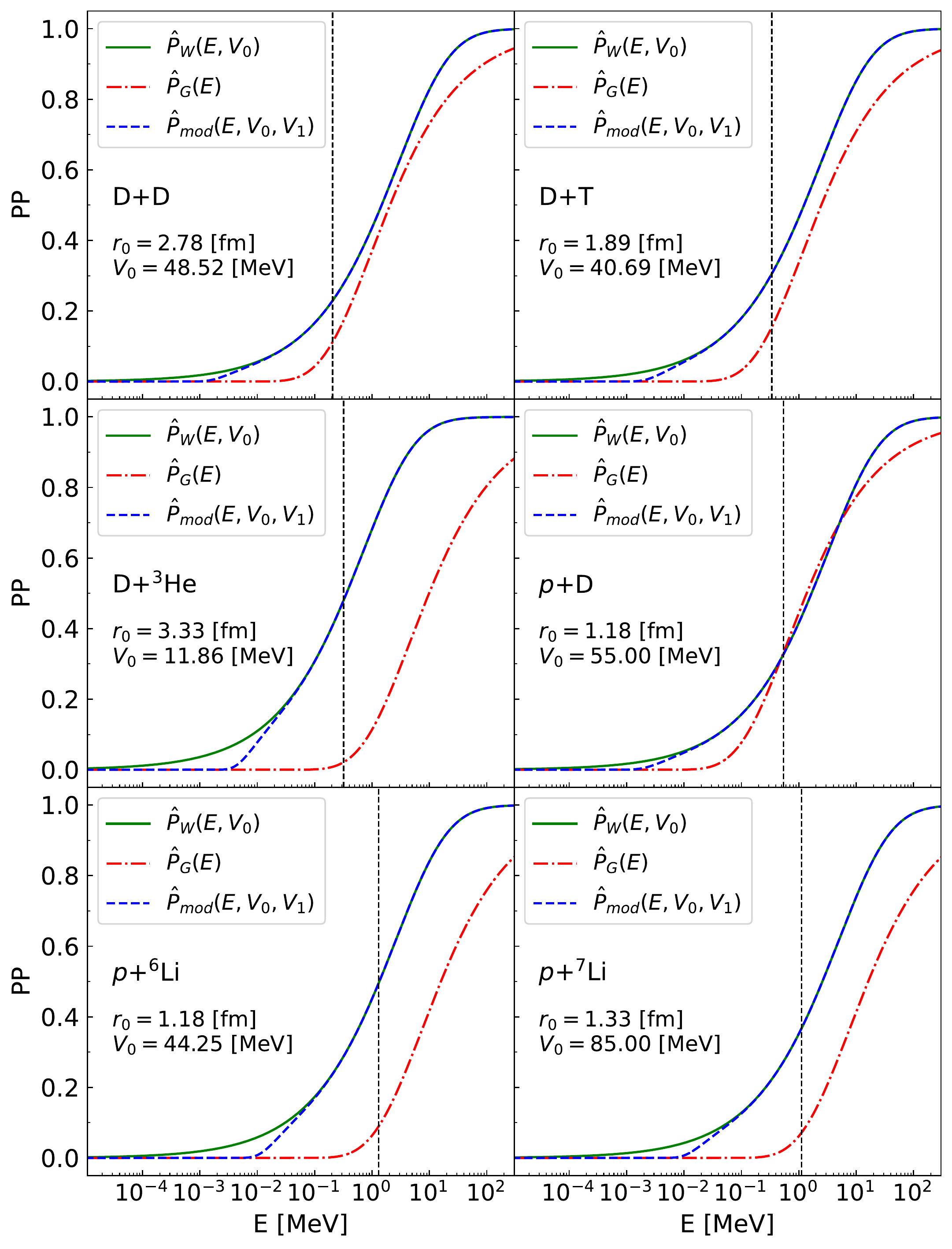}
    \caption{The PP as a function of $E$ for the reactions in Table \ref{tab:couloumb_bariirer}, adopting the fitted $V_0$ and $R_0$ in Ref.\,\cite{2019NuPhA.986...98S}. The green solid, red dashed-dotted, and blue dashed lines indicate the  $\hat{P}_W(E,V_0)$, $\hat{P}_G(E)$, and $\hat{P}_{mod}(E,V_0,V_1)$, respectively. The vertical dashed line marks the Coulomb barrier for each reaction.}
    \label{fig:penetration_probability}
\end{figure}

For the square well potential, parameters of $V_0$ and $r_0$ can be fitted using the complex potential for fusion cross-section data, as reported in Ref.\,\cite{2019NuPhA.986...98S}. The previous literature modeled the square well potential using a complex value, which provided real and imaginary potential depth. Although the complex potential results in a different formula for the penetration probability, the form can be recovered to the formula in Eq.\,(\ref{eq:modified_Gamow_factor}) because the imaginary potential is much smaller than the real potential. We confirm that the given imaginary potential affects the PP by less than 2\%. Therefore, it is valid to use only the real $V_0$ with radius $r_0$ in Ref.\,\cite{2019NuPhA.986...98S}. Taking those parameters for the reactions in Table \ref{tab:couloumb_bariirer}, we show $\hat{P}_{G}(E)$ and $\hat{P}_{mod}(E, V_0,V_1)$ in Fig.\,\ref{fig:penetration_probability} as a function of $E$. Near  the Coulomb barrier, the given $V_0$ results in higher $\hat{P}_{mod}(E, V_0,V_1)$ than $\hat{P}_{G}(E)$, except for the case of ${\rm p+D}$ reaction.

The approximation is valid only for the reaction of ${\rm p+D}$ due to its small reduced mass and Coulomb barrier. Near the Coulomb barrier, since $\hat{P}_{mod}(E, V_0,V_1)$ is approximately equal to $\hat{P}_W(E,V_0)$, the ratio of $\hat{P}_{mod}(E, V_0,V_1)$ to $\hat{P}_{G}(E)$ can be expressed as
\begin{eqnarray}
\frac{\hat{P}_{mod}(E, V_0,V_1)}{\hat{P}_{G}(E)} \simeq \frac{4\sqrt{(E+V_0)E} / (\sqrt{E+V_0} + \sqrt{E})^2}{\exp\left[ -2\pi Z_i Z_j e^2 \sqrt{\frac{\mu_{ij}}{2E}} \right]}.
\label{eq:ratio}
\end{eqnarray}
For the given $V_0$ and condition of $E \simeq V_1$ near the Coulomb barrier, the numerator $\hat{P}_{mod}(E, V_0,V_1)$ in Eq.\,(\ref{eq:ratio}) is a constant and less than unity. Therefore, the ratio in Eq.\,(\ref{eq:ratio}) is determined by the exponential term. Among three reactions of D+D, D+T, and p+D ($Z_i = Z_j = 1$), the p+D reaction has the smallest value of $\sqrt{\mu_{ij}/(2E)}$, resulting in the smallest exponential term. This leads the ratio in Eq.\,(\ref{eq:ratio}) to be close to unity. On the other hand, for the other reactions, $\hat{P}_{mod}(E, V_0,V_1)$ is higher than $\hat{P}_{G}(E)$ near the Coulomb barrier because of the large value of $\sqrt{\mu_{ij}/(2E)}$. This inconsistency implies that the previous assumptions adopted for $\hat{P}_{G}(E)$ oversimplify the PP. 

%
%
\section{MODIFIED GAMOW ENERGY}
For the case of the non-resonant reaction, using the $S(E)$ and $\hat{P}_G(E)$, the thermonuclear reaction rate in Eq.\,(\ref{eq:reaction_rate}) is rewritten as
\begin{eqnarray} \label{eq:reaction_rate2}
    \langle \sigma v \rangle_{ij}
    &=& \sqrt{\frac{8}{\pi \mu_{ij} T^3}} \int S(E) e^{-E/T} \hat{P}_G(E) dE.
\end{eqnarray}
When the low energy region of $S(E)$ is considered, the integrand in Eq.\,(\ref{eq:reaction_rate2}) depends on the Maxwell-Boltzmann (MB) distribution and $\hat{P}_G(E)$. The combination of these two factors results in a sharp peak in energy, referred to as the Gamow peak. This peak represents the energy range where most of the thermonuclear reactions occur. The Gamow peak energy, $E_G$, is determined by the extremum condition in the integrand in Eq.\,(\ref{eq:reaction_rate2}). The expression for $E_G$ is given as $E_G=0.1220\left(Z_i^2 Z_j^2 \mu_{ij}T_9^2 \right)^{1/3} (\text{MeV})$, where $T_9$ is the temperature in GK. This value of $E_G$ can be used to estimate the thermonuclear reaction rate in astrophysical environments or to determine the energy range of interest for measuring the nuclear cross-section in experiments \cite{2007PhRvL..98l2501S,2010PhRvC..81d5807R,2007PhRvC..75d5801N}.

\begin{figure}[h] 
    \centering
    \includegraphics[width=15 cm]{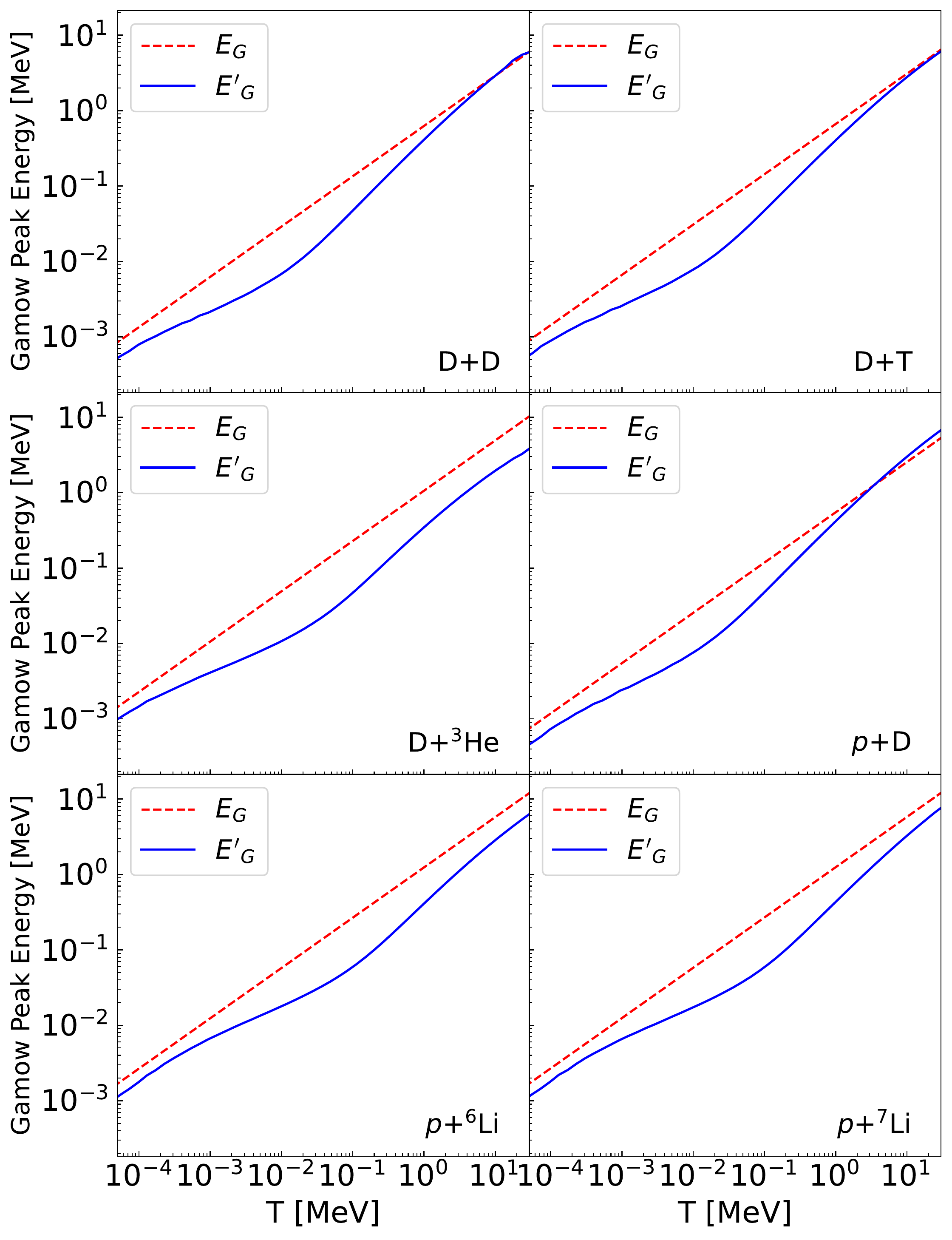}
    \caption{$E'_G$ and $E_G$ as a function of temperature. The blue solid and red dashed lines indicate $E'_G$ and  $E_G$, respectively. }
    \label{fig:Gamow_energy_vs_modified_Gamow_energy}
\end{figure}

Since $\hat{P}_{mod}(E,V_0,V_1)$ differs from the $\hat{P}_G(E)$, the $E_G$ is also changed. Replacing $\hat{P}_G(E)$ to $\hat{P}_{mod}(E,V_0,V_1)$ in Eq.(\ref{eq:reaction_rate2}), the extremum condition is 
\begin{eqnarray}
    \frac{d}{dE}\left(\hat{P}_{mod}(E, V_0,V_1) \exp(-E/T)\right)_{E=E_G'}=0,
    \label{eq:modified_Gamow_energy_condition}
\end{eqnarray}
where $E'_G$ is the modified Gamow peak energy. We note that $E'_G$ depends on $V_0$ and $r_0$ as well as $T$. Adopting the same potential parameters in Fig.\,\ref{fig:penetration_probability}, we show $E_G$ and $E'_G$ as a function of $T$ in Fig.\,\ref{fig:Gamow_energy_vs_modified_Gamow_energy} for the reactions in Table \ref{tab:couloumb_bariirer}. In most temperature regions, $E_G'$s are lower than $E_G$s. 
This is because the increase PP intersects with the MB distribution in the lower energy region. As a result, we find that the $E'_G$ is maximally about 5.3 times smaller than $E_G$ at the sub-keV region. This implies that the estimation of interest energy region in experiments or astrophysical environments could be lower than the case when the $E_G$ is adopted. 

%
%
\section{DISCUSSION: ELECTRON SCREENING EFFECT}
We also discuss the effects of change in PP on electron screening. In a plasma, the Coulomb potential of the charged nuclei is screened by the electron. As a result, the Coulomb potential between $i$ and $j$ is replaced to
\begin{eqnarray}
    V_s(r) = \frac{Z_i Z_j e^2}{r} e^{-r/\lambda_D},
\end{eqnarray}
where subscript $s$ indicates `screened' and $\lambda_D$ is the Debye length. The Debye screening increases the thermonuclear reaction rate by decreasing the Coulomb barrier.

For the conventional Gamow factor, the enhancement factor is defined as
\begin{eqnarray}
f_{g,s}(\rho, T) \equiv \frac{\left\langle \sigma v \right\rangle_{s}}{\left\langle \sigma v \right\rangle_{b}} \simeq \frac{\int e^{-E/T} \hat{P}_{G,s}(E, \rho, T) dE}{\int e^{-E/T} \hat{P}_{G,b}(E) dE},
\end{eqnarray}
where subscripts $b$ indicates `bare'. The PP for bare potential is equal to the Gamow factor, i.e., $\hat{P}_{G,b}(E) = \hat{P}_G(E)$, and the screened PP is given as
\begin{eqnarray}
    \hat{P}_{G,s}(E, \rho, T) &=& \exp \left[ -2 \int^{R_C}_{R_0} \sqrt{2\mu_{ij} \left(\frac{Z_i Z_j e^2}{r}e^{-r/\lambda_D}-E\right)}dr  \right] \\[12pt] \nonumber
    &\approx& \exp \left[ -2 \int^{R_C}_{R_0} \sqrt{2\mu_{ij} \left(\frac{Z_i Z_j e^2}{r}- \frac{Z_i Z_j e^2}{\lambda_D}-E\right)}dr  \right],
    \label{eq:screened_Gamow_factor}
\end{eqnarray}
whose effect on nucleosynthesis has been extensively studied \cite{PhysRevC.83.018801,Mori_2020,2020ApJ...898..163F, 1954AuJPh...7..373S,2013CoPP...53..397P,2021JCAP...11..017H,1997ApJ...488..507I,2002ApJ...579..380I}. On the other hand, for $\hat{P}_{mod}(E,V_0,V_1)$, we can write the enhancement factor as follows:
\begin{eqnarray}
f_{mod,s}(\rho, T, V_0,V_1) \simeq \frac{\int e^{-E/T} \hat{P}_{mod,s}(E,\rho, T,V_0,V_1) dE}{\int e^{-E/T} \hat{P}_{mod}(E,V_0,V_1) dE},
  \label{eq:mod_screened_Gamow_factor}
\end{eqnarray}
where $\hat{P}_{mod,s}(E,\rho,T,V_0,V_1)$ includes the barrier for the Debye potential. Therefore, the enhancement factor in Eq.\,(\ref{eq:mod_screened_Gamow_factor}) depends on $V_0$ and $V_1$ as well as $T$ and $\rho$.

\begin{figure}[h] 
    \centering
    \includegraphics[width=15cm]{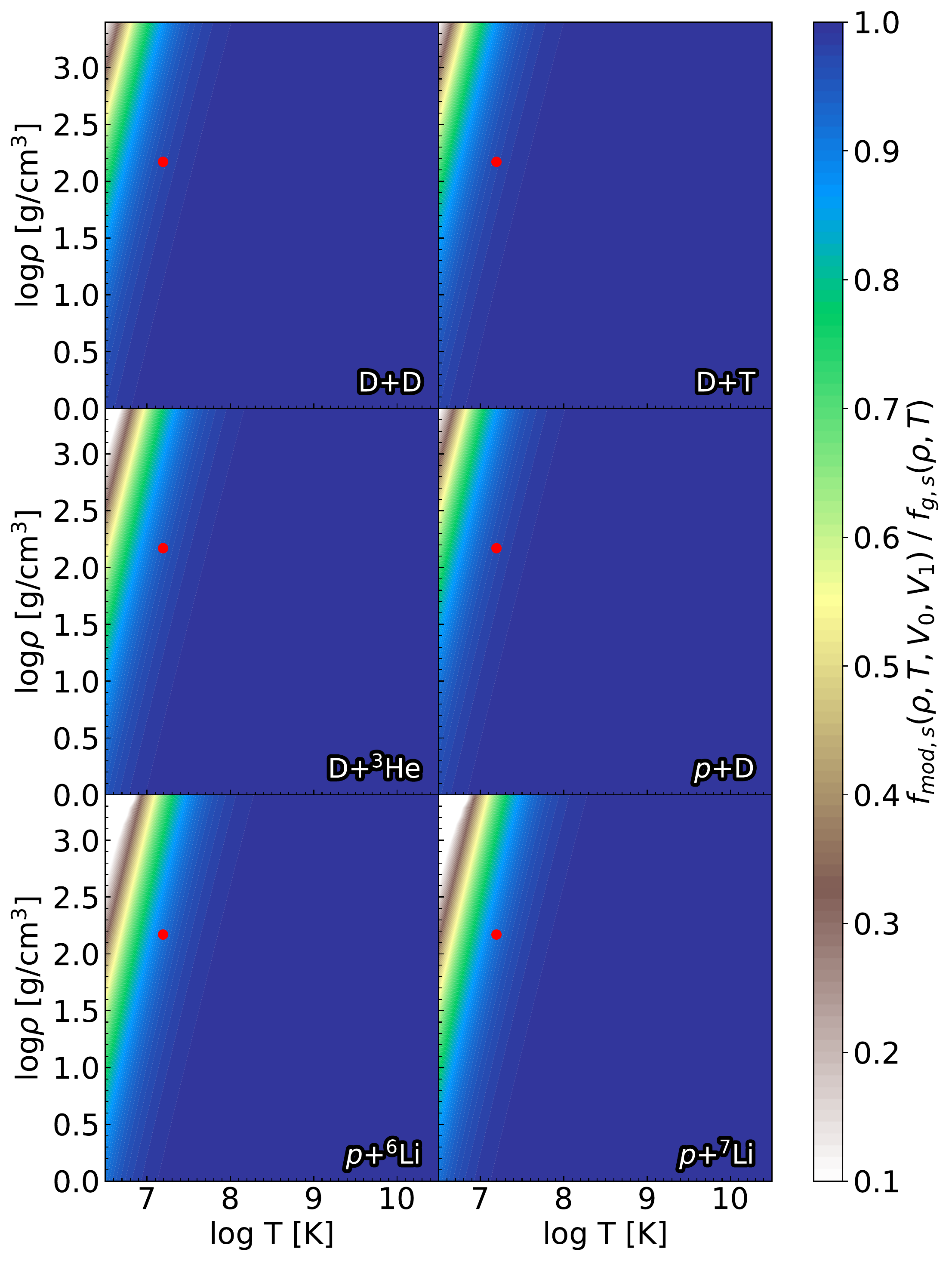}
    \caption{$f_{mod,s}(\rho,T,V_0,V_1)/f_{g,s}(\rho, T)$ as a function of temperature and density. The red dot is marked the solar core condition in Tab. \ref{tab:screen}. }
    \label{fig:enhancement_factor}
\end{figure}

Figure \ref{fig:enhancement_factor} shows the ratio of both enhancement factors, $f_{mod,s}(\rho,T,V_0,V_1)/f_{g,s}(\rho, T)$. Since the effects of plasma is significant at low temperatures and high densities, both enhancement factors are also meaningful in this region. According to Fig.\,\ref{fig:enhancement_factor}, $f_{mod,s}(\rho,T,V_0,V_1)/f_{g,s}(\rho, T)$ is less than unity at the entire parameter space. This is because the increased of PP reduces the Gamow peak energy, the region determining the enhancement factor dominantly. For the central condition of the Sun in the standard solar model \cite{1988RvMP...60..297B}, Table \ref{tab:screen} shows the $f_{mod,s}(\rho_c,T_c, V_0,V_1)/f_{g,s}(\rho_c, T_c)$ where $\rho_c$ and $T_c$ are the central density and temperature of the Sun, as an example result. The difference between $f_{mod,s}(\rho,T,V_0,V_1)$ and $f_{g,s}(\rho_c, T_c)$ implies that the electron screening enhancement factor in the Sun (or other astrophysical environments) could be reconsidered along with the present prescription. We leave it as a future work.

\begin{table}[h]
    \begin{tabular}{c | c | c|c}
        \hline\hline        
        Reactions   &  $f_{g,s}(\rho_c, T_c)$ &  $f_{mod,s}(\rho_c,T_c, V_0,V_1)$ & $~f_{mod,s}(\rho_c, T_c, V_0,V_1)/f_{g,s}(\rho_c, T_c)~$ \\
        \hline     
        D+D         & 1.076      &    1.030     &   0.956 \\ 
        D+T         & 1.076      &    1.030     &   0.956 \\ 
        D+$^3$He    & 1.159      &    1.069     &   0.922 \\ 
        p+D       & 1.077      &    1.025     &   0.952 \\ 
        p+$^6$Li  & 1.249      &    1.094     &   0.877 \\ 
        p+$^7$Li  & 1.249      &    1.097     &   0.879 \\ 
        \hline\hline       
    \end{tabular}
    \caption{ Summary of $f_{g,s}(\rho_c, T_c)$,  $f_{mod,s}(\rho_c, T_c, V_0,V_1)$, and $f_{mod,s}(\rho_c, T_c, V_0,V_1)/f_{g,s}(\rho_c, T_c)$ for the central condition of the Sun, where $T_c = 0.0125~ T_9$ and $\rho_c = 160 ~\rm{g/cm^3}$.}
    \label{tab:screen}
\end{table}

%
%
\section{Conclusion}
In conclusion, in this study, we present a modification to the conventional Gamow factor for the reaction of light nuclei discarding previously adopted assumptions:  1) neglecting the nuclear interaction potential term and 2) assuming that the Coulomb barrier is much higher than the particle energy. Our results indicate that the PP is sensitive to the potential parameter of $V_0$, which was not taken into account in the conventional Gamow factor. For the potential parameter fitted to fusion cross-section data, we also show that the modified PP is higher than the conventional one near the Coulomb barrier. This implies that the previous approximations for the Gamow factor oversimplify the PP for reactions of light nuclei. Furthermore, the increase in PP results in a lower Gamow peak energy, which also reduces the enhancement factor of the thermonuclear reaction rate due to the electron screening effect as summarized in Table \ref{tab:screen}. Therefore, the previous estimation of the interest energy region in experiments or astrophysical environments could decrease, which might be revisited in other studies, such as experiments of reactions of light nuclei, stellar nucleosynthesis, and explosive nucleosynthesis. 

\section*{Acknowledgement}
The work of E.H., H.K., K.H. and M.K.C. are supported by the National Research Foundation of Korea (Grant Nos. NRF-2021R1A6A1A03043957 and NRF2020R1A2C3006177). D.J. was supoorted by the Institute for Basic Science under IBS-R012-D1.
\bibliographystyle{apsrev4-1}
\bibliography{references.bib} 

\end{document}